\documentclass[twoside]{dis07}
\usepackage[latin1]{inputenc}
\usepackage[dvips]{graphicx,epsfig,color}
\usepackage{wrapfig,rotating}
\usepackage{amssymb,amsmath,array}

\begin{document}
\title{Heavy Flavor Production in DGLAP improved Saturation Model}

\author{Sebastian Sapeta$^{1, 2}$
%
%
\vspace{.3cm}\\
%
1- M. Smoluchowski Institute of Physics, Jagellonian University,\\ 
   Reymonta 4, 30-059 Cracow, Poland
\vspace{.1cm}\\
2- Department of Physics, CERN, Theory Division,\\
CH-1211 Geneva 23, Switzerland
}

\maketitle

\begin{abstract}
The saturation model with DGLAP evolution is shown to give good description of the production of the charm and beauty quarks in deep inelastic scattering. 
The modifications of saturation properties caused by the presence of heavy quarks are also discussed.
\end{abstract}

\section{Introduction}

The saturation model of Golec-Biernat and W\"usthoff (GBW)~\cite{gbw} has been very successful in describing both the inclusive, $F_2$, and diffractive, $F^{D(3)}_2$,  structure functions of proton at low values of the Bjorken variable $x$. It incorporates the idea of parton saturation in a simple way by introducing the $x$-dependent saturation scale in the form $Q_s^2(x) = (x_0/x)^{\lambda}$ with the parameters $x_0$ and $\lambda$ determined from the fit to $F_2$. 
The model was then improved by Bartles, Golec-Biernat and Kowalski (BGK)~\cite{Bartels:2002cj} by including the DGLAP evolution of the gluon density, whose effects are important for the small-$r$ part of the dipole cross section. This allowed to describe the new, more precise, HERA data.
However, only the three light quarks contributions to the inclusive structure function $F_2$ were considered by these authors. 

In order to consistently describe DIS one should take into account also the contributions of the heavy quarks since, as found by H1~\cite{h1heavy} and ZEUS~\cite{zeusheavy}, they may reach 30\% for charm and 3\% for beauty.
In this short note~\cite{url} we present the results of the studies~\cite{Golec-Biernat:2006ba} of the DGLAP improved saturation model where also the charm and beauty quarks are present. The parameters of the model with the five flavors are fixed by the fit to the $F_2$ experimental data. Then, the contributions of the heavy quarks to the proton structure function, $F^{c\bar c}_2$ and $F^{b\bar b}_2$, as well as the longitudinal structure function $F_L$ and the diffractive structure function $F^{D(3)}_2$ are \emph{predicted}. 

This study is related to that presented in~\cite{kowalski} where the proton profile function is taken in a Gaussian form. In our case, however, the gluons are assumed to be evenly distributed over a certain area with a sharp boundary. Furthermore, in our approach we do not need to introduce an arbitrary mass for the light quarks as it was done in~\cite{gbw, Bartels:2002cj, kowalski, Iancu:2003ge, Soyez:2007kg}.

\section{The DGLAP improved saturation model}

The dipole picture of the photon-proton interaction at low $x$ has been demonstrated to be a~useful tool for calculating proton structure functions, both inclusive and diffractive. In this framework $\gamma^* p$ interaction is regarded as a two-stages process. Firstly, the photon with the virtuality $Q^2$ dissociates into a quark-antiquark par of a given flavor. This pair, referred to as a color dipole, is characterized by the fraction of the photon momentum carried by the quark (or antiquark) $z$ and the $q\bar q$ transverse distance vector $\vec{r}$. The splitting is described by the photon wave function $\Psi(\vec{r},z,Q^2,m^2_f, e_f)$, quantity fully calculable in quantum electrodynamics (QED). 
In the second stage, described by the dipole cross section $\hat\sigma (x, \vec{r})$,  the color dipole interacts with the proton and, since the non-perturbative contributions are expected to be important, modeling of  $\hat\sigma (x, \vec{r})$ cannot be avoided. The expression for the inclusive proton structure function $F_2$ may be, quite generally, written as

\begin{equation}
F_2(x, Q^2)
= \frac{Q^2}{4\pi^2\, \alpha_{\rm em}}\,
 \sum_{f}\, \sum_{P} \int \!d\,^2\vec{r}\! \int_0^1 \!dz\;
  \vert \Psi_{P}^f\,(\vec{r},z,Q^2,m^2_f, e_f) \vert ^2 \:
  \hat{\sigma}\,(x,\vec{r}),
\end{equation}
where the sums run over the quark flavors $f$ and the polarizations of the virtual photon $P$. 

In the BGK model the following form of the dipole cross section is used
\begin{equation}
\hat{\sigma}(x,r)  = \sigma_0 \left\{1-\exp\left(-
                     \frac{ \pi^2}{3\, \sigma_0}\, r^2\, \alpha_s(\mu^2)\,
                           xg(x,\mu^2)
                     \right)\right\},
\end{equation}
where $\mu^2 = C/r^2 + \mu_0^2$.
It interpolates between the GBW cross section~\cite{gbw} (at large $r$) and the perturbative result~\cite{Frankfurt:1996ri} (at small $r$). Thus, both the feature of color transparency and gluon saturation are incorporated in this definition. The gluon distribution evolves with $\mu^2$ according to the leading order DGLAP equation, simplified further by neglecting quarks,  with the MRST inspired initial condition

\begin{equation}
\label{eq:gluon_int}
xg(x,Q^2_0) =  A_g \, x^{\lambda_g}(1-x)^{5.6}
\qquad {\rm at} \qquad Q^2_0 = 1\ {\rm GeV}^2.
\end{equation}
Altogether, the model has five parameters $\sigma_0$, $C$, $\mu^2_0$, $A_g$ and $\lambda_g$, which are determined by the fit to the $F_2$ data.  The fit with the charm and beauty contributions was performed using the recent data from H1~\cite{Adloff:epjc21} and ZEUS~\cite{Chekanov:epjc21}. The H1 data points were multiplied by 1.05 to account for slightly different normalizations between the two experiments. Since the dipole picture is expected to be valid only at sufficiently small $x$ we restricted ourselves to \mbox{$x<0.01$}. Due to the fact that the gluon density is evolved according to DGLAP equations the model is supposed to work for any value of photon virtuality. Thus, in the fit, we used the entire range of $Q^2$ covered by the data. This gave us 288 experimental points. The light quarks were taken massless and the typical values for the masses of the heavy quarks were used, namely $m_c = 1.3{\rm\ GeV}$ and $m_b = 5.0{\rm\ GeV}$. 
The number of active flavors was set to 4~(for the fit with charm) or 5~(for the fit with charm and beauty), the value of $\Lambda_{\rm QCD}= 300$~MeV, and the argument in the dipole cross section was modified $x \to x\left(1 + 4m_f^2/Q^2\right)$ similarly to~\cite{gbw, Bartels:2002cj}.

\section{Fit results and model predictions}

\begin{table}[t]
\begin{center}
\begin{tabular}{|l||c|c|c|c|c||c|} \hline 
& $\sigma_0\,$[mb] & $A_g$ & $\lambda_g$ & $C$ & $\mu^2_0$ & $\chi^2/$ndf
\\ \hline  \hline
light + c + b                  & 22.7 &~1.23~&~- 0.080~~&~0.35~&~1.60~&~1.16~
\\ \hline
light + c                      & 22.4 &~1.35~&~- 0.079~~&~0.38~&~1.73~&~1.06~
\\ \hline
light                          & 23.8 &13.71 &~~0.41~&~11.10~&~1.00~& 0.97
\\ \hline
\end{tabular}
\end{center}
\caption{The results of our fits with heavy quarks to the $F_2$ data and 
the massless fit from~\cite{Bartels:2002cj}.} 
\label{fit_results}
\end{table}

The results of the fit with heavy quarks are summarized in Table \ref{fit_results}, where also the  light quarks fit parameters from~\cite{Bartels:2002cj} are recalled for reference. We see that the quality of the fit in terms of $\chi^2$/ndf is good. Adding heavy quarks results in a rather dramatic change of the parameters of the model. In particular, the sign of the power $\lambda_g$ alters which means that the initial gluon distribution grows with decreasing $x$ oppose to the  the case of the light quarks fit where it was valencelike.

\begin{figure}[t]
\begin{center}
\includegraphics[width=0.46\columnwidth]{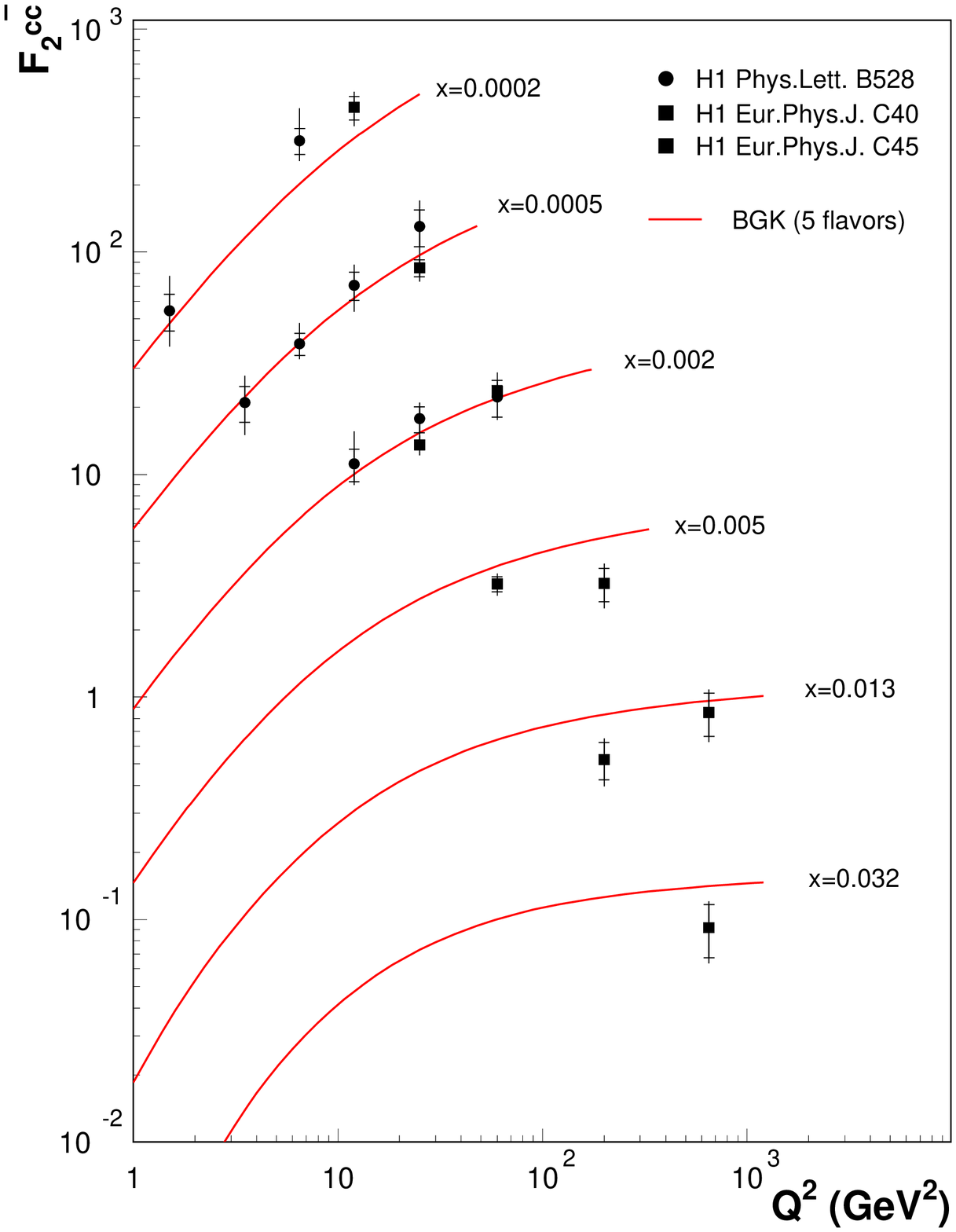}
\hspace{0.3cm}
\includegraphics[width=0.46\columnwidth]{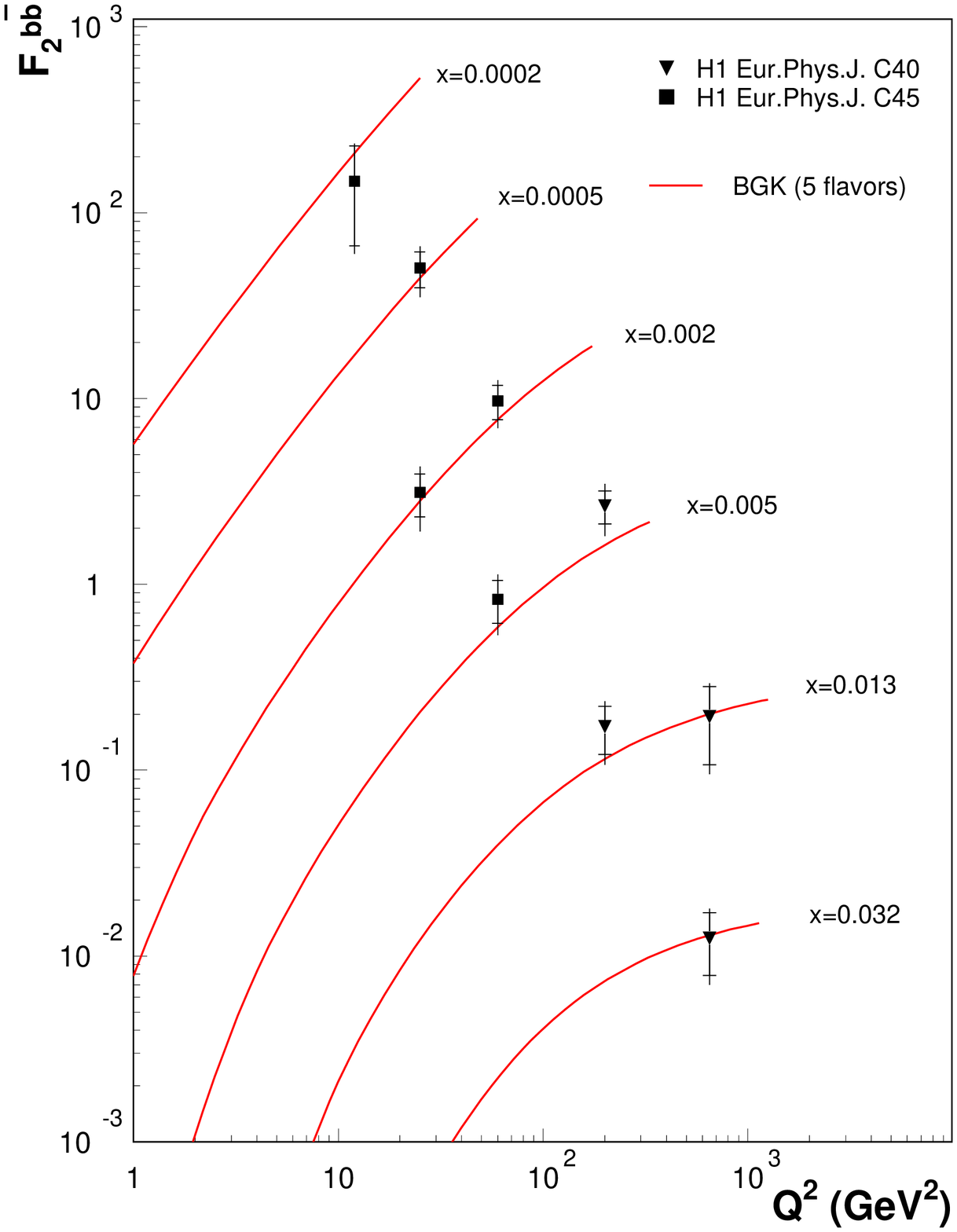}
\caption{
Predictions for the charm ($\times 4^n$) and beauty ($\times 8^n$) structure functions from the BGK model with heavy quarks compared with the H1 data~\cite{h1heavy}.}
\label{Fig:fcfb}
\end{center}
\end{figure}

The predictions for the heavy quark contributions to the structure function, $F^{c\bar c}_2$ and $F^{b\bar b}_2$, are presented in Fig.~\ref{Fig:fcfb}. We observe very good agreement with the data points from H1~\cite{h1heavy}. This persists even for $x>0.01$ that is beyond the range used in the fit to determine the model parameters.  

The diffractive structure function $F^{D(3)}_2$ was also calculated and good agreement with the H1 and ZEUS data was found. Likewise, the longitudinal structure function $F_L$ obtained from our analysis agrees with the H1 estimations. For more details on $F_L$ and $F^{D(3)}_2$ the reader is referred to~\cite{Golec-Biernat:2006ba}.

\begin{wrapfigure}[19]{r}{0.45\columnwidth}
\centerline{\includegraphics[width=0.44\columnwidth]{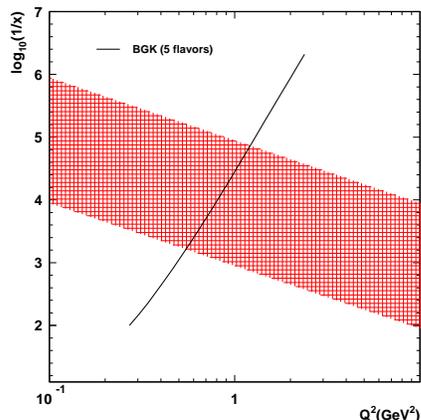}}
\caption{
The critical line from our analysis together with the acceptance region of HERA.}
\label{Fig:critline}
\end{wrapfigure}

Let us finally discuss the effect of heavy quarks on the position of the critical line. This line in $(x,Q^2)$ plane which marks the transition to the saturation region is plotted in Fig.~\ref{Fig:critline}. 
We have checked~\cite{Golec-Biernat:2006ba} that the presence of heavy quarks shifts this line slightly towards the smaller values of $Q^2$ at low $x$. Similar behavior has been already observed in the GBW model~\cite{gbw}. 
Let us also point out that the critical line presented in Fig.~\ref{Fig:critline} is very similar to that obtained by Soyez~\cite{Soyez:2007kg} in the modified Iancu, Itakura and Munier (IMM) model~\cite{Iancu:2003ge} with heavy quarks
(see \cite{Golec-Biernat:2006ba} and \cite{Soyez:2007kg} for the precise, slightly different, definitions of the critical lines).

\section{Conclusions}

We have shown that the DGLAP improved saturation model with heavy quarks provides successful description of a considerable number of quantities measured in DIS.  In particular, it predicts correctly the charm and beauty contributions to the proton structure function. This result is quite remarkable given the simplicity of the framework we use.
This may suggest that the $k_T$ factorization is a more efficient way of approaching DIS at small $x$ (see also~\cite{jung}) or be considered as an argument supporting the idea of saturation at HERA.

\section{Acknowledgments}
I would like to express my gratitude to Krzysztof Golec-Biernat with whom this work has been done.
It is also a pleasure to thank Leszek \mbox{Motyka} for numerous valuable discussions during this workshop.
The support from  Marie Curie ESRT Fellowship of the European Community's
Sixth Framework Programme under contract number (MEST-CT-2005-020238) and
the grant of Polish Ministry of Science N202 048 31/2647
(2006-08) are gratefully acknowledged.


\begin{footnotesize}

\end{footnotesize}


\end{document}